\documentclass[runningheads,a4paper]{llncs}

\usepackage{amssymb}
\usepackage{graphicx}

\usepackage[utf8]{inputenc}

\newcommand{\censor}[1]{{#1}}

\usepackage{url}
\urldef{\mailsa}\path|{alexander.schneider, christian.meter}@hhu.de|    
\newcommand{\keywords}[1]{\par\addvspace\baselineskip
\noindent\keywordname\enspace\ignorespaces#1}

\urldef{\mailsb}\path|{hagemeister, mauve}@cs.uni-duesseldorf.de|    

\begin{document}

\mainmatter  % start of an individual contribution

\title{Tor is not enough: Coercion in Remote Electronic Voting Systems}

\titlerunning{Tor is not enough: Coercion in Remote Electronic Voting Systems}

\author{Christian Meter \and Alexander Schneider \and Philipp Hagemeister\and Martin Mauve}
\authorrunning{Tor is not enough: Coercion in Remote Electronic Voting Systems}

\institute{Heinrich-Heine-University,\\
Universitätsstr. 1, 40225 Düsseldorf, Germany\\
\mailsa\\
\mailsb\\
\url{http://www.cn.hhu.de}}


\toctitle{}
\tocauthor{}
\maketitle

\begin{abstract}
Current electronic voting systems require an anonymous channel during the voting phase to prevent coercion.
Typically, low-latency anonymization-networks like Tor are used for this purpose. In this paper we devise a monitoring 
attack
that allows an attacker to monitor whether participants of an election voted, despite the use of a
low-latency network during the voting phase, thereby breaking an important part of coercion-freeness.
We implement a simulation carrying out our attack and measure its success rates.
\end{abstract}

\section{Introduction}
\label{chapter:introduction}

Electronic voting sees a lot of adoption recently. For example, electronic voting systems are used in national 
elections in Estonia \cite{madise2006voting},
for country council elections in Norway \cite{stenerud2012reality} and for canton-wide votes in Switzerland \cite{gerlach2009three}.\\
To compete with traditional paper ballots, electronic voting systems must ensure that the voter has the same or better
privileges regarding criteria such as anonymity and integrity. 
One important criterion is coercion-freeness, which is defined as follows:\\
\emph{A remote voting system must provide security aspects to prevent an 
	attacker, the coercer, from
	being able to force a voter to behave in a certain way during the election. 
	In a coercion-free system the coercer must be unable to force the voter to disclose her voting credentials, vote for a 
	certain party or abstain from the election.}\\
We focus on the abstaining part of the coercion-freeness definition.
Many remote voting systems rely on mix-nets like Tor to anonymize the identity of the voter during the voting
phase. But adversaries with sufficient resources -- like intelligence agencies -- may have the capability to 
control some parts of the Internet to see the traffic flow of voters and voting servers. With these information an 
attacker is able to deduce whether a voter has cast a ballot.
Using a discrete-event network simulator we demonstrate a network monitoring attack and the deanonymization 
of the voters.\\
In the following section \ref{chapter:relatedwork} we briefly review related work. In section 
\ref{chapter:theory} we describe how correlation attacks can break coercion freeness.\\
Section \ref{chapter:simulation} describes a correlation attack and the setup, implementation and results of our 
simulation.
Section \ref{chapter:realtor} bridges the gap between our simulation and real world systems. The findings are then 
discussed in section \ref{chapter:discussion} and a conclusion follows in section \ref{chapter:conclusion}.

\section{Related Work}
\label{chapter:relatedwork}

Several correlation-based attacks on low-latency anonymity networks are known \cite{levine2004timing,shmatikov2006timing,bauer2007low}. While those
attacks concentrate on attacking flaws of the network itself, our approach does not attack the network 
but instead exploits its latency preserving nature.
Steven J. Murdoch and Piotr Zieli{\'n}ski \cite{murdoch2007sampled} published a paper about how ISP data can be used to
execute traffic analysis on users of Tor. The main idea of the methods is related to our pattern matching approach but
concentrates on general deanonymization of arbitrary users of Tor while we focus specifically on certain users -- the
voters participating in an election.
J. Benaloh \cite{benaloh2013rethinking} published a study reviewing the state of coercion with focus on electronic 
voting and new technologies, showing that coercion is a serious threat to the safety and integrity of electronic 
voting systems.

\section{Attacking Coercion Freeness}
\label{chapter:theory}
In this chapter we want to sketch out the main idea behind a correlation attack.

First, we have to focus on the last part of the definition given in the introduction, the ability to force voters to 
abstain from an election. This might seem insignificant, but in reality can enable a coercer to manipulate an election.
For example consider an election where a coercer has a list of known voters of party \emph{A}. If the coercer forces
those voters to abstain, the results for all other parties automatically improve relative to party \emph{A}. The coercer
can use money, force or other means to oblige a potential voter to obey.\\
The coercer must have the capability to verify if her victims obey and abstain from the election. If a coercer does not have this capability, the victims can disobey without fear of repercussions from the coercer. 
An anonymous channel of some sort during the voting phase should be employed to deny the coercer the ability to 
monitor voter participation.
If it is possible to break coercion-freeness, the voting system must be considered unsafe.\\
Most electronic voting systems designed for large-scale elections in fact assume some kind of anonymous channel -- at least during the voting phase -- without providing one and instead rely on existing low-latency anonymity networks like Tor.\\
In the following we show that such a low-latency network is not enough. Consider a scenario where the attacker 
monitors a subgroup of all voters, because the attacker has access to some ISP backbones and also taps the connection 
before the ballot boxes.
Consider furthermore a typical voter who uses Tor for web-browsing and online voting. At some time during the 
web-browsing the voter decides to cast her vote. The vote is tunneled through Tor, so technically the attacker only 
sees arbitrary packets leaving the voter's machine but does not know if those packets are web-traffic or the casting 
of a vote.\\
The problem is that low-latency networks are mostly latency preserving. An attacker could simply observe the time between the individual packets and watch for a similar pattern on the side of the ballot boxes. Other hints like less web-traffic on the voter's side during the voting process could potentially simplify the correlation attempts of the attacker.
If an attacker is able to deduce that a voter did in fact take part in an election with a reasonably high chance, the property of coercion-freeness is not fulfilled.\\
Imagine a coercer who coerces the known voters of a certain candidate into not voting at all, and does know which ISP the voters use. Now she observes the traffic and uses a correlation attack to determine who voted despite the instructions to abstain. The voter who did not follow the coercer's instructions is completely oblivious to the fact that the coercer knows that she voted.
The simulation of such a pattern-based correlation attack is described in the next section.

\section{Simulation of a Traffic Monitoring Attack}
\label{chapter:simulation}

We now demonstrate that the attack described above is in fact feasible by means of simulation.

\subsection{Attacker Model}
\begin{figure*}
	\includegraphics[width=\textwidth]{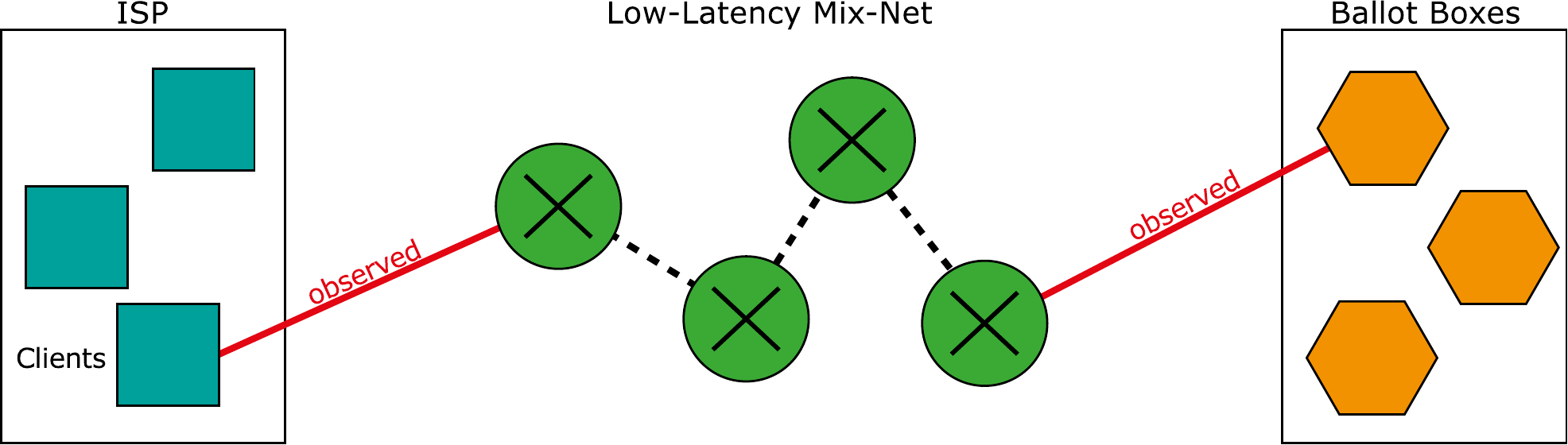}
	\caption{Scheme of Attacker Model}
	\label{fig:attackermod}
\end{figure*}
In our experiment we assume the attacker has access to the following parts: First she has the control over the complete
traffic from one ISP. So even if a client browses the web with an encrypted connection, the attacker is still able to 
see
the IP addresses of each packet being sent.\\

Second she is able to observe the traffic of the ballot boxes. Therefore she can see all packets coming to or
getting out of the ballot boxes. Assuming there are only TLS encrypted connections, the attacker has again at least the
IP addresses of the source and destination of the packets.

We assume the attacker has no further information. In particular, she is not able to watch
the packet flow inside Tor. In conclusion the attacker sees all traffic on the first hop link of the monitored clients 
and on the link leading to the ballot box servers. All other traffic is invisible to the attacker. This is illustrated 
in figure \ref{fig:attackermod}.

These are reasonable assumptions for the real world, because sufficiently large organizations, like intelligence 
agencies, could be able to take control over such channels. 

\subsection{User Model}
We simulate our own network to get all information about the packet flows and to realize our correlation attack. In this
network we run ballot boxes, file and web servers, Tor nodes and many clients browsing the Internet and placing a vote
randomly in one of the ballot boxes. The following subsections describe the different user-types inside the simulation.

\paragraph{Distributed Ballot Boxes}
In a reliable voting system there must be multiple ballot boxes, which allow to vote in equal measure. Because of that 
our system allows multiple ballot boxes to prevent unavailability.\\
Ballot boxes of existing voting systems differ in their traffic pattern but usually communicate with the voter using 
at least three packets\cite{clarkson2008civitas}. We implemented our
ballot box with a one packet reply. This makes a monitoring attack even harder than with three or more packets, 
because there is less data to correlate.

\paragraph{Clients}
The clients in our simulation are reduced to a minimum. They establish connections through Tor to a file server and
download some content. If the client is supposed to vote it connects to a ballot box at a random time, places the vote
and terminates the connection. Placing the vote needs only one packet, which is a simplified assumption, because in 
real voting systems
the data requires more than one packet. Again, more packets sent at nearly the same time create a more distinct 
pattern
and makes the analysis easier because it stands out from the background noise.  To achieve a more realistic behavior of
the client, we started a second simulation with the clients visiting some websites.  In preparation we downloaded the
global top ten websites\footnote{Google.com, Facebook.com, Youtube.com, Yahoo.com, Baidu.com, Wikipedia.org,
	Twitter.com, Qq.com, Taobao.com, Amazon.com} published by \emph{Alexa.com} on 2014-07-28 \cite{alexa}, set them up as
web servers in the simulator and configured the clients to randomly visit one of the servers. Remaining parameters
stayed the same, but the results are slightly better with the simulated web browsers than the assumption all clients are
downloading some 1-5 MB files.

\paragraph{Bulk Clients}
Within the simulation are some more clients downloading content in bulk from the file servers. They generate traffic and
are also using Tor. We use them to create background traffic, making it more difficult to find the correct pattern due 
to high network
activity.

%%%%%

\subsection{Simulation}

\subsubsection{The Simulator}
For the simulation we use the open source discrete-event network simulator \emph{Shadow v1.10.0-dev} \cite{shadow}. This
simulator uses the native Tor releases which can be utilized in plug-ins \cite{jansen2011shadow}. Several plug-ins are
provided by the maintainers to create small simulations using Tor. We developed our own plug-in to include a sample
voting application.

\paragraph{Sample Application}
The sample application consists of two parts: the server where a voter can place her vote and a web client actually
placing the vote. A voter, represented by a client, connects to the server through the simulated Tor network and sends
the minimum packets needed to establish a connection. She places a single packet containing the vote and closes the
connection. Our ballot box accepts the connection, receives the vote, sends one packet back with the confirmation
message and waits for the connection-termination by the voter.

\paragraph{Network Traffic}
The published version of Shadow provides PCAP files to analyze network traffic of each simulated node in our network.
These PCAP files are very useful for deep packet inspection, but we do not need them for our analysis since we limit our
analysis to the timing in a flow of packets. Therefore we modified the source code of Shadow to introduce the 
\emph{PacketInspector} which
logs the source and destination IP-addresses of each packet. Filtering the Shadow log file for PacketInspector entries
leads to a minimal view of the complete packet flow during the whole simulation.

\paragraph{Network Characteristics}
We simulate a Tor network with 195 Tor nodes and three directory servers. Besides Tor-related nodes there are 540
clients randomly downloading content from 100 file servers\footnote{This numbers were chosen due to 
	hardware limitations. Bigger simulations ran out of memory.}. From these 540 clients 200 are randomly chosen following 
the
uniform distribution, to place their vote at a random time to one of the ten ballot boxes. And from all the clients we
assume that we are able to see the traffic of 200 random clients (because we are their simulated ISP). To generate more
traffic noise, we added 210 more clients connecting to the file servers and downloading content in bulk.\\ As 
described above, for the second simulation we added 100 web servers randomly simulating the global top ten websites 
and configured the 540 clients to
visit one of the web servers at a random time.

\subsubsection{Hardware}
As our basic system we use a Debian 7.1 server with 12 Cores of Intel\textsuperscript{\textregistered}
Xeon\textsuperscript{\textregistered} CPU E5-2620 v2 @ 2.10 GHz with 48 GB DDR3 RAM. With this setup the simulation took
about 3:15 hours for one hour of simulation-time.

%%%%%%

\subsubsection{Simulation Runs}
We started multiple simulations with the setup described above, each running for one hour of simulation-time, where the
first 30 minutes are used to initialize the Tor network\footnote{During these 30 minutes the clients are idle.}. First
we configured all clients to download files from random servers and some of them to place a vote randomly. This leads to
busy nodes and a big log file. Analyzing these streams takes a long time, but produces satisfying results.

In the second experiment we reconfigured the clients to connect to simulated websites for a more realistic behavior.
Since connecting to the websites generates less traffic than downloading files, it is even easier to find the voting pattern under these circumstances.

The log files are filtered multiple times. Thereby we made sure, have only the information is available that would be 
available to an attacker. We did this by deleting all
entries not belonging to clients from our simulated ISP.

%%%%%%

\subsection{Analysis of the Simulation Data}
The next step was the analysis of our simulation results. One hour of simulation time created a log of about 60 GB in 
size for the file-transfer user model. Our first step was to reduce the log file and delete all entries that were not 
messages from the PacketInspector. The result was a 20 GB log file solely comprised of entries detailing the 
packet-flow during the simulation.
To simulate our role as an ubiquitous attacker controlling an autonomous system or an ISP we had to somehow limit the traffic we were seeing. So the second step was filtering the remaining entries. We simply acted as if we could only see 200 clients of the simulation and deleted all traffic information that was not going to, or coming from, our visible clients and the voting servers. For the rest of the section we refer to the remaining clients as \textit{visible}.\\

This filtered data is now a mere 500 MB of network traffic. The big task at 
hand is to find the signs in the data that tell us which clients voted when 
and at which voting server and who did not vote at all. To the observer monitoring an ISP backbone, all packets transmitted via Tor or a similar software look fairly alike, because they are encrypted and all have the same size.
The algorithm we are looking for should be defined as follows:\\

\emph{Input:} A list of all visible clients (voters) and servers 
(ballot-boxes). Each client and server is represented by a stream of its in- 
and outgoing packets.\\

\emph{Output:} A list of clients that have voted including time of vote and 
ballot-box used to cast the vote.\\

A perfect algorithm would produce an output-list which would contain precisely all of the clients that voted and none more.
Let us furthermore define the metrics we use to measure the success of the algorithm:\\

\emph{Hit-rate:} Proportion of output-entries that would be also present in 
the output of the perfect algorithm in percent.\\

\emph{False positive:} A false positive is an output-entry that would not be 
present in the output of the perfect algorithm. In other words, the algorithm 
either guessed wrong that a client voted or guessed right but inferred either 
a wrong voting-time or ballot-box.\\
First tries to analyze the traffic were performed using various brute-force 
and timing techniques which produced bad results with far too many false 
positives.
A pattern matching based approach provided better results. The analyzer uses a 
pattern consisting of a sequence that details in which stream the algorithm 
has to look for packets belonging to a vote-connection. 
For example consider a simple ping protocol where a client sends a packet which is received by the server and echoed back.
The sequence \textit{[Client output, Server input, Server output, Client 
	input]} would be the pattern if one would search for events of the type 
described in this example.
This approach allows the analysis to work with arbitrary voting applications where a pattern is known.\\
Acquiring a pattern is usually not a problem, because remote voting software mostly is and should be open source. An attacker can set up a toy example with one server and one client and then use tools like \textit{wireshark} \cite{wireshark}
to extract the corresponding pattern. Scripts for automatic pattern extraction are therefore a real possibility as well.
To show this we wrote a script which takes the log file of a simulation 
containing one voter-client and one voting-server. The voter-client only runs 
the voting-application and Tor. Our script now scans the log file for the 
parts where the server first and last receives communications. Everything in 
between is part of the pattern. Automatic trimming of noise, not belonging to 
a possible pattern, yields the pattern.\\
Where before out of the 200 visible clients between 80 and 100 were false positives, the pattern matching approach 
produced about 50. 50 out of 200 being still far too many we employed additional methods to reduce the false 
positives. 
The problem is that even during one hour a client generates several thousand packets and therefore the pattern matches 
just because for a lot of points in time there is at least one packet present. These several packets can be the Tor
handshakes establishing the circuit or just normal traffic when browsing the web.  In experience most current voting
systems use only few packets to cast a vote. Therefore it is an useful assumption  that when a client sends or receives
more than \textit{x} packets in one second, the traffic most likely is generated by file transfers, video streaming or
similar high traffic services. A noise reduction cuts away all blocks that contain more than \textit{x} packets in a
given time interval \textit{t}.\\
With an optimally chosen \textit{x} this cuts the false positives down to two. In figure \ref{fig1} one can observe the
impact of different values chosen for \textit{x}. While cutting the false positives down to two, we still retained a
hit-rate of 46 out of the 82 for us visible voters.
That means that the algorithm (representing the coercer) is absolutely sure for 56\% of all clients that had cast a 
vote, that they voted and furthermore at which ballot-box and at which time they did so. For the rest the coercer has 
no data that justifies absolute confidence. 
Knowing the time and ballot-box used to vote strengthens the coercers ability to be absolutely sure whether a particular
voter really voted. A random coin-flip would possibly retain a hit-rate of about 50\%, but also have a lot of false
positives and would not provide data the coercer could place confidence in.
The results are similar for the simulation, which used an web-browser-based user model, see figure \ref{fig:browser}.\\
Another parameter is the time interval \textit{d} which describes the time that can at most elapse between every two
packets in the pattern. If at any point in the search for the pattern two consecutive packets are more than \textit{d}
seconds apart, they are discarded as not matching the pattern. Adjustment of \textit{d} can be observed in figure
\ref{fig2} and boosts the hit-rate to 63\% while retaining only two false positives.\\

\subsection{Results of the Analysis}
% % Was sind die Ergebnisse und Implikationen der Simulation?
\begin{figure*}[!ht]
	\centering
	\begin{minipage}{0.49\linewidth}
		\includegraphics[width=\textwidth]{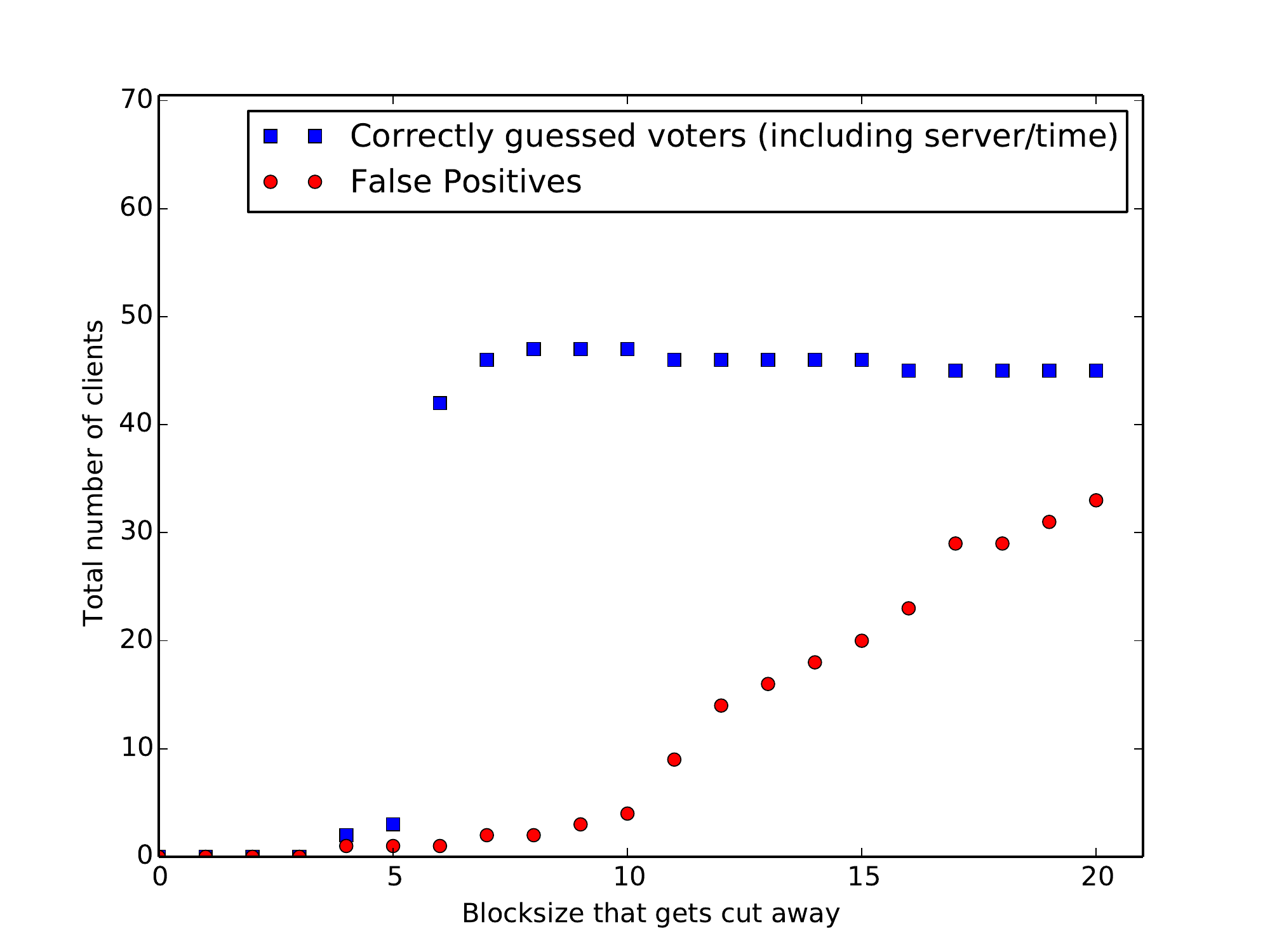}
		\caption{Change of x (maximum number of packets per second which are not discarded from the analysis) with fixed d = 1.0 in the file-transfer simulation}
		\label{fig1}
	\end{minipage}
	\begin{minipage}{0.49\linewidth}
		\includegraphics[width=\textwidth]{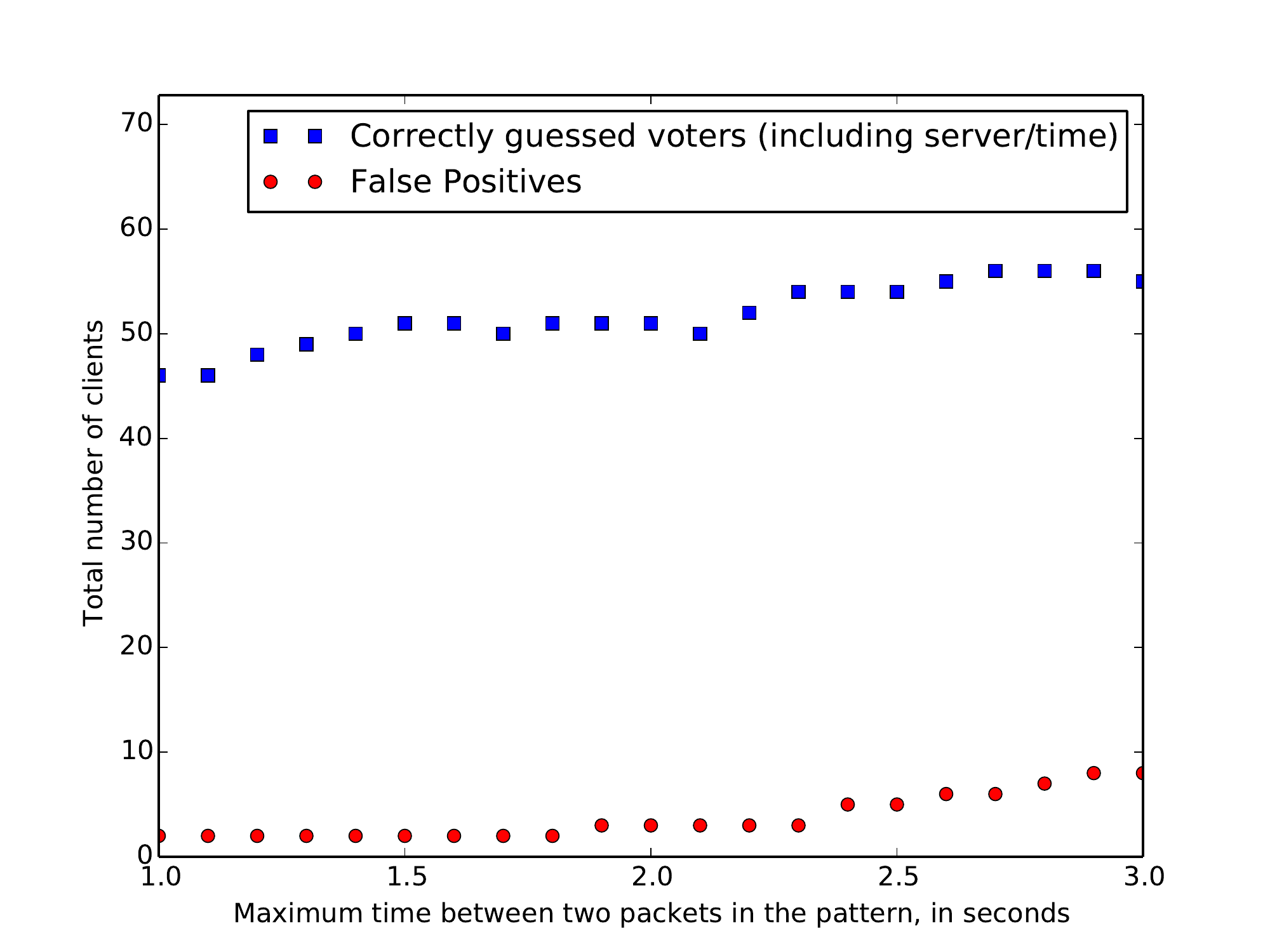}
		\caption{Change of d (Maximum delay between to consecutive packets in a pattern) with fixed block size x = 7 in the file-transfer simulation}
		\label{fig2}
	\end{minipage}
	\begin{minipage}{0.49\linewidth}
		\includegraphics[width=\textwidth]{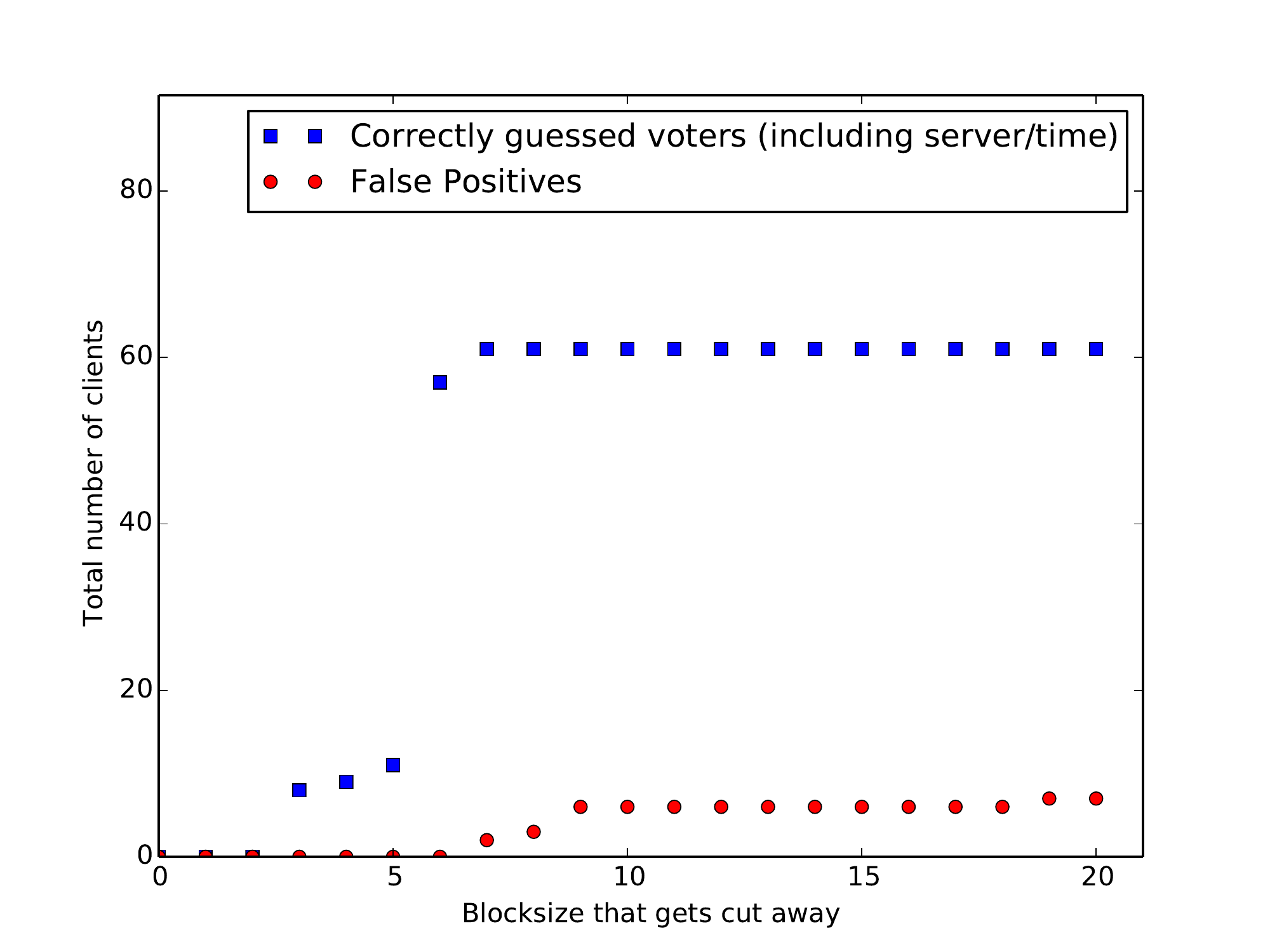}
		\caption{Change of x (maximum number of packets per second which are not discarded from the analysis) with fixed d = 1.0 in the browser-model simulation}
		\label{fig:browser}
	\end{minipage}
	\begin{minipage}{0.49\linewidth}
		\includegraphics[width=\textwidth]{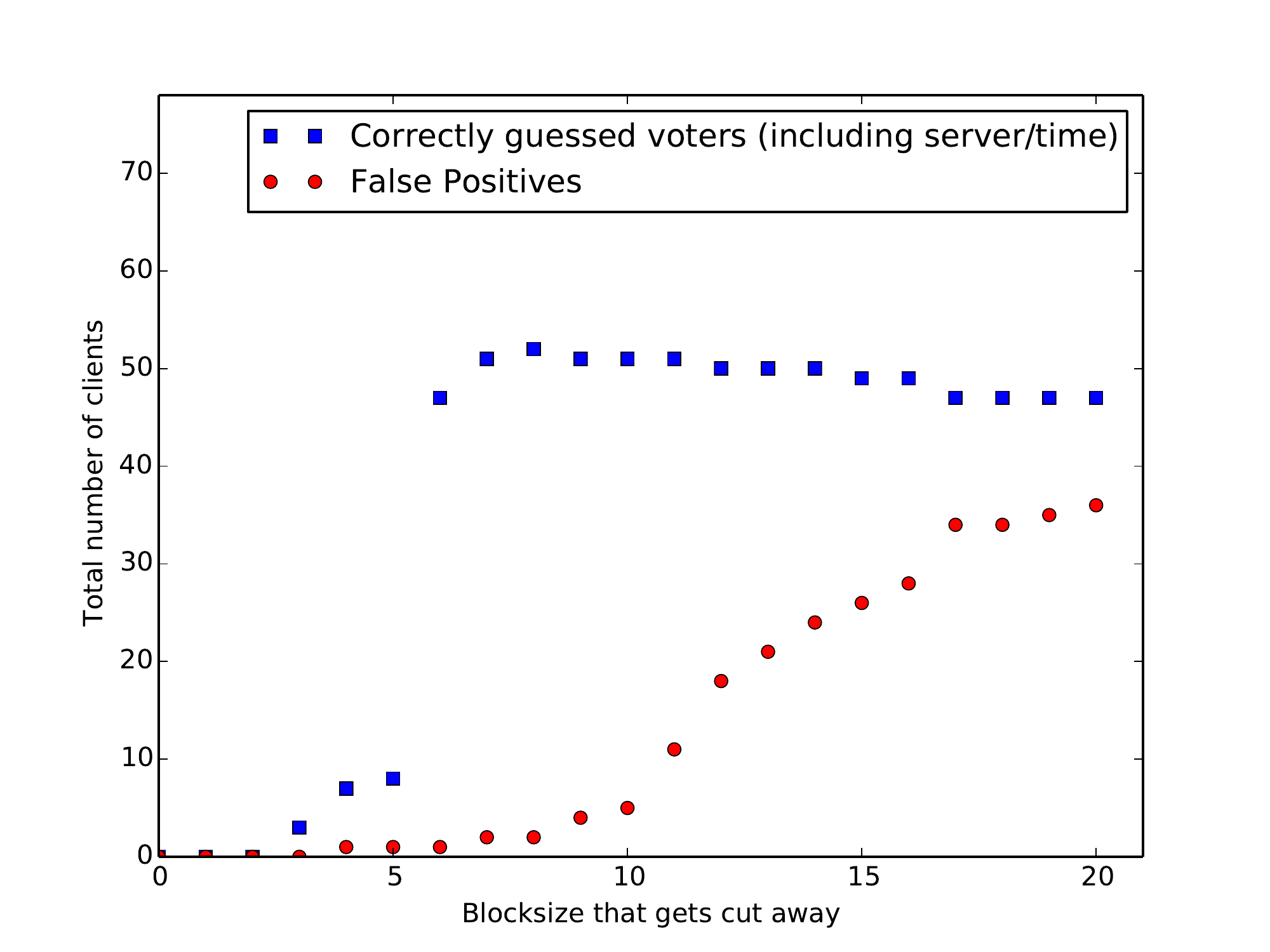}
		\caption{Change of x (maximum number of packets per second which are not discarded from the analysis) with fixed d = 1.5 in the file-transfer simulation}
		\label{fig3}
	\end{minipage}
\end{figure*}

The analysis shows that -- at least in a simulated environment -- it is entirely possible for an attacker with
ubiquitous monitoring capabilities to determine for a large portion of the voters, up to 75\% in the browser-client
simulation, if they cast a vote. These results practically enable large scale coercion, because it is effectively 
possible to monitor the voting behavior for many voters at once. A coercer can then specifically observe if his 
victims abstain from an election like they were ordered to.

We argue that similar results can be obtained in reality. Because most people do not
use software like Tor in their everyday Internet activities it would be even easier to determine if someone has voted
because Tor would only be used for casting the vote and therefore the pattern the attacker is looking for would be more
clear and practically "noise-less". To test this we set up a simulation where the user-clients had Tor running but only
used it to cast their vote. The hit rate for this simulation was about 95\%. This implicates that the property of
coercion-freeness is violated as long as the remote voting software employs low-latency anonymity networks as it's
anonymous channel during the vote. The gap between simulation and real world is bridged further in section 
\ref{chapter:realtor}.\\
The results were obtained conducting two example simulation runs (one with the file-transfer model and one with the 
browser model). Because a simulation run is very consuming regarding time and computational power\footnote{One 
	simulation run occupies our server for about seven hours, not including the analysis of the results}, we did not 
perform a statistical analysis over a significant amount of simulation runs (e.g. 100 runs for each model).

As one can see in figures \ref{fig1}, \ref{fig2}, \ref{fig3} the success of a pattern-based analysis depends on two
parameters. The figures \ref{fig1}, \ref{fig3} illustrate how different grades of noise reduction influence the quality
of the results. If the maximum block size is five or smaller, the results hardly produce any matches. This is to be
expected given the pattern in this case, which is five packets long: Unless the communication spreads over multiple
(one-second) intervals, our analysis prunes even plain vote traffic (without any noise) at such a small maximum block
size. One can furthermore clearly see a peak of hits at a block size of seven and a steady decline of hits thereafter.
This can be explained through the fact that with bigger packet-blocks more noise is present and therefore more false
positives are found just because there are enough packets that would potentially match the pattern, but are in fact not
packets representing the vote.\\
Figure \ref{fig2} shows the influence of the maximum allowed time, in seconds, between two packets in a pattern. The
higher this time, the more hits are possible. The reason is simple: the bigger the allowed delay the higher the chance
that at least one packet falls inside the period and therefore matches the pattern. Unfortunately the false positives
start to rise after a certain size of the delay as well, so that the overall-quality of the results sinks.

The results obtained using the Internet-browser user model were largely the same and even slightly better, as seen in
figure \ref{fig:browser}. This result makes sense, because the loading of a web-site creates less packets than the
loading of several files up to 5 MB, and therefore less background-noise is present to distract the matching algorithm.

\section{Real World Implications}
\label{chapter:realtor}
This section provides a bridge to close the gap between simulation and real-world.
For comparison we used a TCP dump of a real Tor node operated by the Computer Networks Group in the \censor{University of
	Düsseldorf}. With this traffic we are able to show that our pattern matching does
not randomly match the background-noise produced by other clients. Furthermore we studied the network pattern of 
a real world remote voting system to see if our simulation is realistic.

\subsection{Pattern Matching in Real-World Traffic}
We took one hour of the measured traffic of a real Tor-node for this analysis. Each IP address communicating with the 
node is interpreted
to be a client trying to connect to one of the ballot boxes. This traffic represents the network activity an attacker
could see, if she controls an ISP. We assume this is the traffic of the first hop between the client and an entry-node
of Tor.\\
We then added the complete network streams of our ballot boxes from the simulation to this traffic to simulate a 
possible voting scenario.
So the complete traffic contains the communication of the first hop and the packets coming from and going to our ballot
boxes.

Now we have a complete stream as it was generated in the simulation and we can begin to analyze it as we did in section
\ref{chapter:simulation}.

Our pattern matching analyzer tried to match if one of the 1742 possible "voters" had cast a vote, which they had not
since all connections belong to a real-world Tor node not participating in our experiment. The algorithm produced seven
false positives which is a rate of about 0.4\%. 

This experiment shows that the pattern matching approach is not distracted by background-noise of real-world Tor
traffic.

\subsection{Real Remote Voting Systems}
To validate how realistic our simulation is, we examined the Civitas voting system \cite{clarkson2008civitas}. The
implementation was released 2008 as a coercion-free remote voting system. We examined a minimal setup with five voters
and the required infrastructure for Civiats in place. Wireshark was used to inspect the network flow of the experiment.
The packet flow shows that Civitas uses three TCP packets to deliver the vote in contrast to our simulated voting
application which uses only one packet for the vote. In total it uses four packets more than our application (two
extra packets for the payload and two for the corresponding acknowledgements).

We argue that the simulation is thus realistic. The bigger a pattern is the easier it gets to separate it from
background noise. The other way around, finding a very small pattern in a lot of noise, is harder\footnote{If the pattern gets bigger the relative amount of noise decreases}.

\section{Discussion}
\label{chapter:discussion}

In this chapter we discuss the findings of the previous chapters and their implications for the design and
implementation of remote voting systems. \\ It should be clear that using no means for voter anonymity during
communication with any of the election related authorities and servers is not a good idea. Coercion-freeness would be
automatically forfeit if there is an adversary with the power to monitor the connections to and from the election
authorities / servers because a coercer can determine if a voter was abstinent from the election as the coercer wished.
This is not an unrealistic adversary because the existing systems mostly use either a centralized architecture or a
marginally distributed one as in the case of the ballot boxes of Civitas.\\
Obviously some kind of anonymous channel is needed and also required by a lot of the current remote voting systems. 
The question one may ask here is how to realize this anonymous channel. As far as we know, there are no systems which 
implement an anonymous channel themselves. Some popular recommendations are low-latency mix-networks like Tor 
\cite{dingledine2004tor} or Tarzan \cite{freedman2002introducing}. The problem with low-latency networks is that an 
adversary with certain powers can deanonymize the network-traffic, as shown in \cite{levine2004timing,shmatikov2006timing,overlier2006locating,murdoch2007sampled,bauer2007low}.\\
As we have shown through our experiments, low-latency networks do not even have to be attacked directly if the 
attacker can monitor a sufficient portion of all voters. As long as the network is latency-preserving a possible 
correlation outside of the anonymity network can be performed.\\ Adding more packets to hide a possible pattern only 
helps the pattern finding algorithm, because it is easier to find bigger patterns.
We therefore suggest to abstain of the use of low-latency mix-networks for the purpose of creating an anonymous 
channel for remote elections for the above mentioned reasons.\\
Since low-latency networks cannot reliably provide an anonymous channel for remote elections, one has to turn and search
for alternatives. High latency networks seem to be immune to the vulnerabilities presented in the paragraph above.
Mixminion, Freenet or Nonesuch \cite{danezis2003mixminion,clarke2001freenet,heydt2006nonesuch} are examples for high-latency networks.
While all of them provide unlinkability like the low-latency equivalents, Nonesuch additionally provides sender 
anonymity. To send something through Nonesuch, one has to post a steganographically encoded picture to a Usenet user 
group. This way an observer does not know, if one is participating in a user group or using the anonymity network.\\
While high-latency networks may possibly guarantee an anonymous channel, the user voting experience may suffer. A vote
would travel for four or more hours through this network before it would reach its destination. An user could in the
meantime not be sure, that the vote was sent correctly, or even forget to check on the vote all together. Missing
feedback could lead to frustration with the system.\\
Another possible solution could be random faulty packets sent by the voter client. If the client randomly sends 
packets where 
one of the encryption layers for Tor is encrypted with an invalid key, it would cause the discarding of the packet by 
one of the Tor nodes on the path. We are not sure if this solution would be sufficient, because it may also be 
possible 
to adapt the pattern matching algorithm to deal with missing packets inside a stream.

\section{Conclusion}
\label{chapter:conclusion}

In this paper we focused on a monitoring attack carried out by an attacker with ubiquitous Internet surveillance 
capabilities. After constructing the attack theoretically we carried out 
different simulations. The results show that with the help of the monitoring attacks we can determine of a large 
portion of voters whether they voted or not, despite the use of a low-latency mix-net by the voting-software as an 
anonymous link. This practically breaks coercion-freeness. We used different real world data, to show that our 
simulations provide realistic results.\\
An electronic voting system that is not coercion-free is considered unsafe, and 
therefore all systems that do not provide an anonymous channel themselves and instead confide in low-latency anonymity 
networks should be considered unsafe.\\
We conclude that systems which do not provide coercion-freeness because of the previously stated flaws should not be 
used in important real-world and large-scale elections, like presidential elections. An exception are elections in a 
low-coercion environment like student parliament elections, where those systems can still be used.\\
Possible future work would be to examine if a correlation is still possible if the voting servers are implemented as a 
\emph{hidden service} in Tor or when they are simultaneously acting as relay-nodes and voting servers. In the first 
case it would probably be a lot harder because the physical and virtual location of the server is unknown and in 
the second case a lot more cover traffic looking similar to the votes would be provided.\\
Further research if high-latency networks really solve the problem and if they even provide realistic usability should 
also be conducted. The search for an anonymous channel that can be employed in remote electronic voting should be a 
priority to guarantee coercion-freeness.

\bibliographystyle{splncs}
\bibliography{survey}

\begin{thebibliography}{10}
\providecommand{\url}[1]{\texttt{#1}}
\providecommand{\urlprefix}{URL }

\bibitem{alexa}
Alexa~Internet, I.: Alexa - actionable analytics for the web.
  \url{http://www.alexa.com} (Jul 2014)

\bibitem{bauer2007low}
Bauer, K., McCoy, D., Grunwald, D., Kohno, T., Sicker, D.: Low-resource routing
  attacks against tor. In: Proceedings of the 2007 ACM workshop on Privacy in
  electronic society. pp. 11--20. ACM (2007)

\bibitem{benaloh2013rethinking}
Benaloh, J.: Rethinking voter coercion: The realities imposed by technology.
  Presented as part of the USENIX Journal of Election and Technology and
  Systems (JETS) pp. 82--87 (2013)

\bibitem{clarke2001freenet}
Clarke, I., Sandberg, O., Wiley, B., Hong, T.W.: Freenet: A distributed
  anonymous information storage and retrieval system. In: Designing Privacy
  Enhancing Technologies. pp. 46--66. Springer (2001)

\bibitem{clarkson2008civitas}
Clarkson, M.R., Chong, S.N., Myers, A.C.: Civitas: Toward a secure voting
  system. Institute of Electrical and Electronics Engineers (2008)

\bibitem{danezis2003mixminion}
Danezis, G., Dingledine, R., Mathewson, N.: Mixminion: Design of a type iii
  anonymous remailer protocol. In: Security and Privacy, 2003. Proceedings.
  2003 Symposium on. pp. 2--15. IEEE (2003)

\bibitem{dingledine2004tor}
Dingledine, R., Mathewson, N., Syverson, P.: Tor: The second-generation onion
  router. Tech. rep., DTIC Document (2004)

\bibitem{wireshark}
Foundation, W.: Wireshark software homepage. \url{http://www.wireshark.org/}
  (Jul 2014)

\bibitem{freedman2002introducing}
Freedman, M.J., Sit, E., Cates, J., Morris, R.: Introducing tarzan, a
  peer-to-peer anonymizing network layer. In: Peer-to-Peer Systems, pp.
  121--129. Springer (2002)

\bibitem{gerlach2009three}
Gerlach, J., Gasser, U.: Three case studies from switzerland: E-voting. Berkman
  Center Research Publication No  3,  2009 (2009)

\bibitem{heydt2006nonesuch}
Heydt-Benjamin, T.S., Serjantov, A., Defend, B.: Nonesuch: a mix network with
  sender unobservability. In: Proceedings of the 5th ACM workshop on Privacy in
  electronic society. pp. 1--8. ACM (2006)

\bibitem{shadow}
Jansen, R.: The shadow simulator. \url{https://shadow.github.io} (Jul 2014)

\bibitem{jansen2011shadow}
Jansen, R., Hooper, N.: Shadow: Running tor in a box for accurate and efficient
  experimentation. Tech. rep., DTIC Document (2011)

\bibitem{levine2004timing}
Levine, B.N., Reiter, M.K., Wang, C., Wright, M.: Timing attacks in low-latency
  mix systems. In: Financial Cryptography. pp. 251--265. Springer (2004)

\bibitem{madise2006voting}
Madise, {\"U}., Martens, T.: E-voting in estonia 2005. the first practice of
  country-wide binding internet voting in the world. Electronic voting  86
  (2006)

\bibitem{murdoch2007sampled}
Murdoch, S.J., Zieli{\'n}ski, P.: Sampled traffic analysis by
  internet-exchange-level adversaries. In: Privacy Enhancing Technologies. pp.
  167--183. Springer (2007)

\bibitem{overlier2006locating}
Overlier, L., Syverson, P.: Locating hidden servers. In: Security and Privacy,
  2006 IEEE Symposium on. pp. 15--pp. IEEE (2006)

\bibitem{shmatikov2006timing}
Shmatikov, V., Wang, M.H.: Timing analysis in low-latency mix networks: Attacks
  and defenses. In: Computer Security--ESORICS 2006, pp. 18--33. Springer
  (2006)

\bibitem{stenerud2012reality}
Stenerud, I.S.G., Bull, C.: When reality comes knocking norwegian experiences
  with verifiable electronic voting. Electronic Voting  205,  21--33 (2012)

\end{thebibliography}

\end{document}